\documentclass[sigconf,nonacm,balance=false]{acmart}
\usepackage{booktabs}
\usepackage{array}
\usepackage{caption}
\usepackage{xurl}
\newcommand{\allPdpBase}{36.7}
\newcommand{\allPdpWith}{39.3}
\newcommand{\allRecBase}{13.4}
\newcommand{\allRecWith}{14.5}
\newcommand{\allTwo}{5.2}
\newcommand{\allTwoBase}{26.2}
\newcommand{\allTwoCI}{[3.0, 7.4]}
\newcommand{\allTwoWith}{31.4}
\newcommand{\exUnlinked}{81}
\newcommand{\ftPriceNoPriceGap}{0.93}
\newcommand{\ftPriceNoPriceGapCI}{[0.19, 1.63]}
\newcommand{\gapStoreMention}{2.7}
\newcommand{\gapStoreMentionCI}{[2.0, 3.5]}
\newcommand{\genericPdpBase}{55.2}
\newcommand{\humAdj}{1.9}
\newcommand{\humBuyAsMention}{29}
\newcommand{\humCantTell}{7}
\newcommand{\humDecoyN}{24}
\newcommand{\humDecoyRate}{4}
\newcommand{\humMatchItems}{160}
\newcommand{\humMatchPlacN}{39}
\newcommand{\humMatchPostN}{90}
\newcommand{\humPrecPlac}{54}
\newcommand{\humPrecPlacCI}{[39, 68]}
\newcommand{\humPrecPost}{93}
\newcommand{\humPrecPostCI}{[86, 97]}
\newcommand{\humSpecAcc}{75}
\newcommand{\humSpecAccCI}{[65, 82]}
\newcommand{\humSpecItems}{102}
\newcommand{\lnkExtraNew}{42.5}
\newcommand{\lnkExtraRecent}{16.8}
\newcommand{\lnkGap}{14.7}
\newcommand{\lnkGapCI}{[11.9, 17.6]}
\newcommand{\lnkNewFirst}{27}
\newcommand{\lnkRecentFirst}{12}
\newcommand{\medianMinSpec}{19}
\newcommand{\mmNamedDid}{3.1}
\newcommand{\mmNamedDidCI}{[2.5, 3.6]}
\newcommand{\mmNamedPlac}{0.05}
\newcommand{\mmNamedPost}{0.04}
\newcommand{\mmVisitDid}{7.8}
\newcommand{\mmVisitDidCI}{[6.8, 9.0]}
\newcommand{\mmVisitNullAbs}{0.8}
\newcommand{\namedNotPerson}{1.9}
\newcommand{\namedNotPersonCI}{[1.5, 2.2]}
\newcommand{\namedSpec}{2.0}
\newcommand{\namedSpecCI}{[1.7, 2.4]}
\newcommand{\namedSpecPlac}{0.02}
\newcommand{\namedSpecPost}{2.07}
\newcommand{\nvoOtherGap}{0.1}
\newcommand{\nvoOtherGapCI}{[\textminus{}2.4, 2.5]}
\newcommand{\originPerson}{8}
\newcommand{\pairedShareSpec}{53}
\newcommand{\pbNamedWeek}{14}
\newcommand{\pbOrdWeek}{19}
\newcommand{\popHeadSpec}{5.1}
\newcommand{\popPoolSpec}{10.5}
\newcommand{\qArrPlac}{1}
\newcommand{\qArrPost}{28}
\newcommand{\qProdPost}{9.3}
\newcommand{\qProdRatio}{24}
\newcommand{\refMatchPlac}{64}
\newcommand{\refMatchPost}{96}
\newcommand{\refSpecAcc}{83}
\newcommand{\repAllPdp}{1.6}
\newcommand{\repAllPdpCI}{[\textminus{}0.4, 3.4]}
\newcommand{\repAllRec}{2.0}
\newcommand{\repAllRecCI}{[0.5, 3.4]}
\newcommand{\repAllTwo}{4.3}
\newcommand{\repAllTwoCI}{[2.8, 6.1]}
\newcommand{\repMmNamedDid}{3.0}
\newcommand{\repMmNamedDidCI}{[2.5, 3.4]}
\newcommand{\repMmNamedPlac}{0.03}
\newcommand{\repMmNamedPost}{0.00}
\newcommand{\repMmNamedShare}{58}
\newcommand{\repNamedSpec}{2.2}
\newcommand{\repNamedSpecCI}{[1.9, 2.6]}
\newcommand{\repPbNamedWeek}{11}
\newcommand{\repPbOrdWeek}{19}
\newcommand{\repShufReal}{42}
\newcommand{\repShufShuf}{1.2}
\newcommand{\repUnnLo}{1.6}
\newcommand{\repUnnLoCI}{[1.2, 1.9]}
\newcommand{\rtDirectPdp}{23}
\newcommand{\rtDirectShare}{18}
\newcommand{\rtLinkPdp}{83}
\newcommand{\rtLinkShare}{11}
\newcommand{\rtLinkUsual}{2}
\newcommand{\rtOtherPdp}{26}
\newcommand{\rtOtherShare}{17}
\newcommand{\rtSearchPdp}{45}
\newcommand{\rtSearchPerKAfter}{110}
\newcommand{\rtSearchPerKUsual}{56}
\newcommand{\rtSearchShare}{54}
\newcommand{\rtSearchUsual}{55}
\newcommand{\rtTypedPdp}{25}
\newcommand{\rtUnlinked}{89}
\newcommand{\rtpNamedLand}{34}
\newcommand{\rtpNamedNav}{56}
\newcommand{\rtpOtherLand}{18}
\newcommand{\rtpOtherNav}{51}
\newcommand{\searchDay}{+3}
\newcommand{\searchDayCI}{[\textminus{}6, 11]}
\newcommand{\searchDayLoAbs}{6}
\newcommand{\shufReal}{44}
\newcommand{\shufShuf}{1.5}
\newcommand{\ssPdpBase}{20.5}
\newcommand{\stabChatGPT}{2.2}
\newcommand{\stabChatGPTCI}{[1.8, 2.7]}
\newcommand{\stabFirstHalf}{2.1}
\newcommand{\stabGemini}{1.0}
\newcommand{\stabGeminiCI}{[0.5, 1.6]}
\newcommand{\stabNoLargest}{2.2}
\newcommand{\stabPanelist}{2.3}
\newcommand{\stabProlific}{2.4}
\newcommand{\stabSecondHalf}{2.0}
\newcommand{\unnLo}{1.1}
\newcommand{\unnLoCI}{[0.8, 1.5]}
\newcommand{\unnPlac}{1.4}
\newcommand{\unnPostHi}{3.5}
\newcommand{\unnPostLo}{2.6}
\newcommand{\visitSpec}{5.1}
\newcommand{\visitSpecRatio}{2.8}

\newcommand{\infobox}[2]{\begin{figure}[tb]\Description{#1}\noindent{\setlength{\fboxsep}{7pt}\fcolorbox{black!25}{black!4}{\parbox{\dimexpr\columnwidth-2\fboxsep-2\fboxrule}{\small\setlength{\leftmargini}{1em}\setlength{\parskip}{3pt}#2}}}\end{figure}}
\setcopyright{none}
\renewcommand\footnotetextcopyrightpermission[1]{}
\begin{document}

\title[Straight to the Product]{Straight to the Product:\texorpdfstring{\\}{ }What Online Retailers Miss When Shoppers Ask AI First}

\author{Michael Iannelli}
\orcid{0000-0002-5967-2026}
\affiliation{%
  \institution{Scrunch AI}
  \city{New York}
  \country{USA}}
\email{michael@scrunchai.com}

\author{Alan Ai}
\orcid{0009-0004-8467-1633}
\affiliation{%
  \institution{Scrunch AI}
  \city{New York}
  \country{USA}}
\email{alan.ai@scrunchai.com}
\renewcommand{\shortauthors}{Iannelli and Ai}

\begin{abstract}
When people ask an AI assistant what to buy, what happens next at the store?
We link the same people's assistant conversations to their later browsing at twenty large online retailers.
People go to the product the assistant named: among shoppers who had not visited the store in the past week, visits to the named product within a day rise from \namedSpecPlac{} to \namedSpecPost{} per 100 answers (+\namedSpec{} points \namedSpecCI), and on days with a product-page visit at the store, \shufReal\% show a product from their own answer, against \shufShuf\% for a matched answer given to someone else.
The result replicates in a separate five-month period.
Few arrive through the assistant's link: after recommending answers, \rtUnlinked\% of visits carry no sign of it, and \rtSearchShare\% follow a web search, the same share as their usual visits. When a search comes just before the first visit, it names the answer's product \qArrPost\% of the time (\qArrPlac\% in earlier weeks), as if the search box were an address bar.
The chat names the product; the store logs a search. Once there, shoppers keep browsing.

We propose \emph{product follow-through}, the share of answers followed by a visit to the product they named, as a way to evaluate shopping assistants and agents by what people do next: it needs no referral data or provider logs and can be computed for each assistant.
\end{abstract}



\begin{teaserfigure}
\centering
\includegraphics[width=\textwidth]{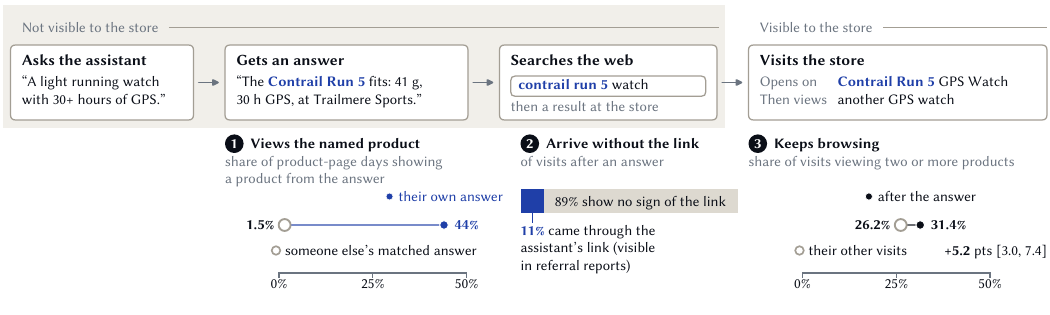}
\caption{One shopping journey, invented for illustration (the watch and the store are fictional), with the paper's three findings. The assistant names a product; many shoppers reach the store through search and keep browsing. The store sees only the last step.}
\Description{Four cards joined by arrows: ask the assistant, get an answer, search the web, visit the store. The first three are shaded as not visible to the store. Below, three small charts: the product page matches the shopper's own answer far more often than someone else's; a bar of 100 visits in which a thin slice comes through the assistant's link; and visits viewing two or more products, more often than on other visits.}
\label{fig:teaser}
\vspace{6pt}
\end{teaserfigure}

\maketitle

\begin{figure*}[t]
\centering
\includegraphics[width=\textwidth]{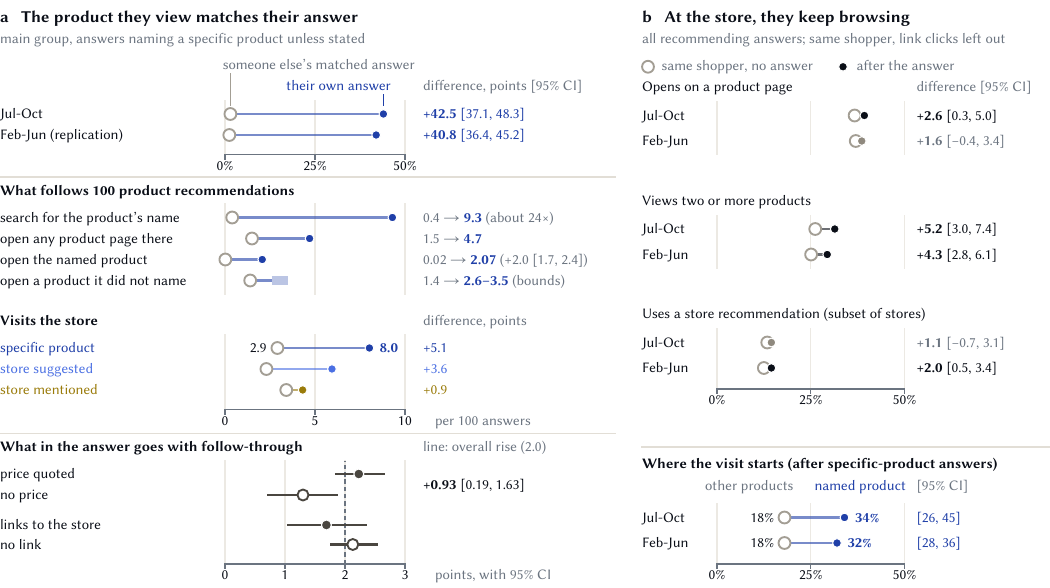}
\caption{(a) People view the product from their own answer, not someone else's, in both periods; the more specific the step, the larger the ratio to the baseline; recommendations move store visits far more than mentions; a quoted price goes with a larger rise. (b) At the store, visits open on a product page only slightly more often than usual, shoppers browse a little more, and the named product often opens the visit. Rings and dots are the groups labeled in each block; the pale bar for products the answer did not name spans the lower and upper bounds.}
\Description{Panel a: two dumbbells from about one percent to over forty percent for own versus someone else's answer, in two periods; below, store visits and named-product visits by answer type, rising most for specific-product answers. Below that, a per-100 row for viewing a product the answer did not name, drawn as a range between bounds, and the answer features that go with follow-through. Panel b: dumbbells for opening on a product page, viewing two or more products and using store recommendations, each with a small gap, and the share of visits that start on the named product against other products.}
\label{fig:main}
\end{figure*}

\section{Introduction}
A shopper asks an assistant for a light running watch with at least 30 hours of GPS.
The answer names one, the Contrail Run 5, and a store that sells it.
A few minutes later the shopper types ``contrail run 5'' into a search engine, clicks a result and lands on the watch's page at the store.
The store's analytics record a search visit to a product page.
The conversation that named the watch leaves no trace there (Figure~\ref{fig:teaser}).

How common is this journey, and what does it mean for retailers?
Earlier panel work found that brands an assistant recommends get more searches and site visits~\cite{iannelli2026fptp}.
Here we go down to the product.
We link the same people's assistant answers to their browsing at twenty large retailers and follow the journey in order:

\begin{enumerate}
\item \textbf{People go to the product the assistant named.} On \shufReal\% of days with a product-page visit at the named store, they view a product from their own answer, against \shufShuf\% for someone else's matched answer.
\item \textbf{The store rarely sees the assistant.} \rtUnlinked\% of visits after a recommending answer show no link from it. Most follow a web search, as usual visits do, often for the product's name.
\item \textbf{Then they keep browsing.} Against all their other visits, they view two or more products more often, in both periods, and otherwise shop much as usual: how often a visit opens on a product page or uses the store's recommendations changes by a few points at most (Figure~\ref{fig:main}).
\end{enumerate}

A search for the product's name follows \qProdPost\% of specific-product answers, about \qProdRatio{} times the usual rate, and the store's analytics log the visit as ordinary search traffic.

The study makes two contributions.
It gives product-level evidence, across sites and without referral data, that an assistant's answer shapes which product people go to see.
It also turns the answer-to-product connection into a measure, product follow-through (Section~\ref{sec:named}), which shows retailers and assistant builders what referral reports cannot.

\paragraph{Related work.}
Most shopping-assistant evaluations stop at the answer.
Conversational recommenders are studied for how they elicit preferences and present options~\cite{zhang2018conversational,jannach2021survey}, LLM-based ones with simulated users~\cite{friedman2023recllm}, and shopping agents in simulated stores~\cite{yao2022webshop} or against expert rubrics~\cite{fan2026shoppingreasoning}.
Work on generative engine optimization asks how products get named in answers~\cite{aggarwal2024geo,bagga2025egeo}.
None of these follows the shopper out of the conversation.

Studies of what happens next see one side of the traffic.
Kaiser and Schulze measure retailer sessions that carry a ChatGPT referral~\cite{kaiser2026chatgpt}; Pew finds that people click search results less often when an AI summary appears~\cite{chapekis2025pew}; usage data from one travel platform show chat and search interleaved in the same purchase journey~\cite{yan2026platform}; and a controlled experiment compares searching with chatting~\cite{xu2023chatgpt}.
Panel studies link assistant use to later brand visits~\cite{iannelli2026fptp} and to the order of searching and asking~\cite{iannelli2026shape}.
We go to the product level and follow the same people across sites, from the answer to the product page, with or without a referral.

\section{Data and Measurement}\label{sec:data}
\paragraph{Panel and answers.} An opt-in consumer panel in two English-speaking markets records the web pages people visit and, for the same people, their conversations with ChatGPT, Gemini, Claude, Perplexity and Grok.\footnote{We do not report panel size, per-analysis user counts or per-cell sample sizes; these are commercially sensitive. Effect sizes, rates and intervals are reported throughout, and every main estimate uses the full eligible panel, not a sample. All intervals are 95\%; the label-validation intervals are Wilson intervals, and all others come from a bootstrap that resamples people.}
The main period runs from July to October 2026, and February to June 2026 is a replication: the plan for its core analyses was committed before its data were pulled, and the later at-store analyses reuse the same code.
For each person and each of twenty large retailers, we take the first answer that names the store.
A language-model judge sorts the answers: an answer names a \emph{specific product} there, like the Contrail Run 5, \emph{suggests the store} for the person's need, or only \emph{mentions} it.
The first two are \emph{recommending answers}.
The \emph{main group} keeps answers in which the assistant, not the person, first named the store, for people who had not been to that store in the past week.

\paragraph{Comparisons.} Every shopper is measured against themselves.
The 24 hours after an answer are set against the same person's visits to that store 14, 21 and 28 days earlier (averaged: the \emph{earlier-week baseline}).
For the product match, each answer is also paired with someone else's answer that names the same store, in the same month, for the same kind of request, country and product category; the judge deciding whether a visited page shows the answer's product is not told whose answer it is.
And because people who chat may simply be busy online~\cite{lewis2011here}, we also take matched moments: the same person and store, the same three-hour time of day, weekday or weekend, recent activity and past-week visit status, at least three days away and with no answer naming any of the twenty stores within a day.
Around those moments store visits do not rise; they dip by \mmVisitNullAbs{} points. Paired answer by answer (the \pairedShareSpec\% of specific-product answers with a matched moment), store visits after a specific-product answer still rise \mmVisitDid{} points \mmVisitDidCI{} more than around the moment (a more active subset than the +\visitSpec{} in Figure~\ref{fig:main}a).

Each comparison addresses a different alternative explanation: the earlier weeks account for a person's habits at a store, the swapped answer for a store's popular products and the category, and the matched moments hold fixed how busy the person was online and when.

\paragraph{Matching answers to products.} An answer names a store when its text matches one of the twenty retailers' names. For each such answer we keep the person's last message before it and the parts of the answer around the store's first two mentions. Both judges are classification tasks run on one underlying model, which picks among written options, each with a criterion, instead of writing free text. The answer-type judge reads the request and the answer excerpts. The product-match judge reads the same excerpts and the product pages the person then visits at that store, each described by its page title and web address. It counts a match when a page shows a product the answer named, allowing a different size, color or pack or another listing of the same item. A page of the same kind of product, or a page it cannot judge, is not a match. The steps need only answer text and later page views, so they carry over to other assistants and panels.

\paragraph{Label validation.} Two other frontier models re-label samples from every judge; where they agree with each other, they confirm \refMatchPost\% of the product-match judge's matches after an answer and \refMatchPlac\% in the earlier-week baseline, and agree with the answer-type judge on \refSpecAcc\% of answers.
A reader blind to the judge's labels and to the time window audited \humMatchItems{} product-match items, \humCantTell{} of which it could not judge: the judge's matches were confirmed in \humPrecPost\% \humPrecPostCI{} of \humMatchPostN{} after an answer and \humPrecPlac\% \humPrecPlacCI{} of \humMatchPlacN{} in the earlier-week baseline, where true matches are rare, and called \humDecoyRate\% of \humDecoyN{} page sets the judge had rejected a match; adjusting for these precisions (missed matches, which the full answer text would add, are not corrected) leaves the named-product rise at \humAdj{} points against \namedSpec{} unadjusted.
On a class-balanced sample of \humSpecItems{} answers, the same reader agreed with the answer-type judge \humSpecAcc\% \humSpecAccCI{} of the time. Its errors run one way: the reader confirmed every mention, but called \humBuyAsMention\% of the judge's recommending answers mentions, which would shrink the recommending rises, not inflate them.

\section{Findings}\label{sec:findings}

\subsection{People go to the product the assistant named}\label{sec:named}
Take the days on which a main-group shopper opened a product page at the store a specific-product answer named: on \shufReal\% of them, a page shows a product from the shopper's own answer.
Swap in someone else's matched answer and the share drops to \shufShuf\% (Figure~\ref{fig:main}a).
The result replicates in a separate five-month period: \repShufReal\% against \repShufShuf\%.

Counted over all such answers, whether or not a product page followed, \namedSpecPost\% lead to a visit to the named product within a day, against \namedSpecPlac\% in the earlier-week baseline: a rise of \namedSpec{} points \namedSpecCI, and \repNamedSpec{} \repNamedSpecCI{} in the earlier period.
Every step of the journey rises over the baseline (Figure~\ref{fig:main}a, middle), and both kinds of recommendation move store visits far more than a passing mention does.
The more specific the step, the larger the ratio to the baseline: store visits rise \visitSpecRatio{} times, searches for the product's name about \qProdRatio{} times, and visits to the named product's page about a hundred times, from \namedSpecPlac{} to \namedSpecPost{} per 100 answers. Recommending goes with far more store visits than mentioning: a store suggestion adds \gapStoreMention{} points \gapStoreMentionCI{} more than a bare mention, which adds under one visit per 100 answers.
Only \originPerson\% of matched products had been named by the shopper earlier in the conversation, and without them the rise barely moves (\namedNotPerson{} points \namedNotPersonCI).

Four checks leave this pattern in place: someone else's answer, the earlier period, a blind human reader, and moments with no answer. Around those moments, with pages scored against the same answer, visits to its product do not rise: \mmNamedPost{} against \mmNamedPlac{} per 100 answers (earlier period \repMmNamedPost{} against \repMmNamedPlac). For the \pairedShareSpec\% of specific-product answers with a moment (\repMmNamedShare\% earlier), the rise after the answer is \mmNamedDid{} points \mmNamedDidCI{} larger than around it, paired answer by answer (earlier period \repMmNamedDid{} \repMmNamedDidCI). That subset is more active than the overall +\namedSpec{} (+\repNamedSpec{} earlier). Nor is it driven by a few heavy users or one big store: the named-product rise is \stabPanelist{} points with each person counted once, \stabProlific{} without the 5\% of shoppers with the most answers, \stabNoLargest{} without the largest retailer, and \stabFirstHalf{} and \stabSecondHalf{} in the two halves of the period.

We call this rate \emph{product follow-through}, defined in the box.
It can be reported separately for each assistant: follow-through rises \stabChatGPT{} points \stabChatGPTCI{} after ChatGPT answers and \stabGemini{} \stabGeminiCI{} after Gemini answers.
It also differs with what the answer contains (Figure~\ref{fig:main}a, bottom). Answers that quote a price near the store's name are followed by a larger rise than those that do not (+\ftPriceNoPriceGap{} points \ftPriceNoPriceGapCI), while naming one product or several, or linking to the store, makes no detectable difference in named-product visits.

\infobox{A boxed definition of product follow-through: what is counted, what counts as a match, the comparison, the data needed, and a worked example.}{\textbf{Product follow-through.} The share of answers naming a specific product that are followed, within 24 hours, by a visit to that product's page at the named store.
\begin{itemize}\setlength{\itemsep}{1pt}
\item \emph{Counted over:} every such answer, whether or not the person visits the store.
\item \emph{A match:} the visited page shows a product the answer named; size, color and pack variants count.
\item \emph{Compared with:} the same person's visits to that store 14, 21 and 28 days earlier.
\item \emph{Needs:} the answer text and the same person's later page views; no referral link, provider logs or integration with the assistant.
\end{itemize}
\emph{Worked example.} The Contrail Run 5 answer counts once in the denominator. The shopper's visit to the watch's page a few minutes later counts as a match. Across main-group answers naming a specific product, \namedSpecPost\% follow through, against \namedSpecPlac\% in earlier weeks. On the narrower set of days with a product-page visit at the store, \shufReal\% show the answer's product, against \shufShuf\% for someone else's answer.}

\subsection{The store rarely sees the assistant}\label{sec:routes}
\begin{figure}[t]
\centering
\includegraphics[width=\columnwidth]{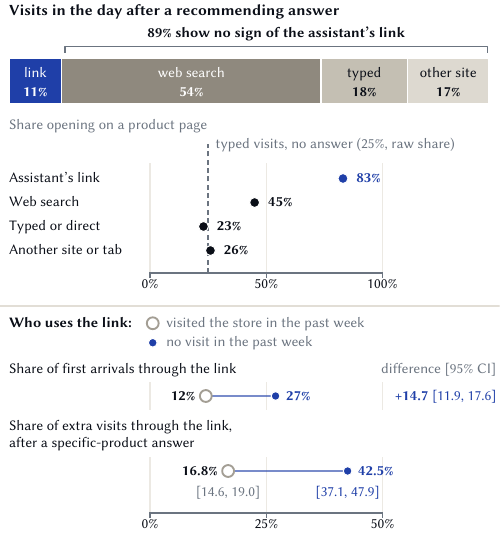}
\caption{Most visits after a recommending answer arrive by web search; only a thin slice comes through the assistant's link, and the link is used more by shoppers who had not visited the store in the past week. Raw shares of visits, February to October; the bottom block counts first arrivals and extra visits above each person's earlier weeks.}
\Description{A stacked bar of 100 visits: assistant's link \rtLinkShare, web search \rtSearchShare, typed or direct \rtDirectShare, another site or reopened tab \rtOtherShare. Below, the share opening on a product page by route: link \rtLinkPdp\%, search \rtSearchPdp\%, direct \rtDirectPdp\%, other \rtOtherPdp\%, with a reference line at the typed-visit rate of \rtTypedPdp\%.}
\label{fig:routes}
\end{figure}
Our shopper did not click the assistant's link, and most visits show no sign of one. A visit comes through the link when it carries an AI referrer or an AI marker in its address, and follows a web search when it has a search referral, a search ad or a search in the 30 minutes before; the remaining visits are typed or direct, or come from another site or a reopened tab, checked in that order.
Of 100 visits after a recommending answer, \rtLinkShare{} come through the link (\rtLinkUsual{} in their usual visits); the rest show no sign of it, and \rtSearchShare{} follow a web search (Figure~\ref{fig:routes}), as \rtSearchUsual\% of their usual visits do. Search visits double, from \rtSearchPerKUsual{} to \rtSearchPerKAfter{} per 1,000 answers, so the extra traffic looks like ordinary search traffic.
Counting only the extra visits on each route (visits in the day after, minus the same person's earlier-week average) changes little: \exUnlinked\% still show no sign of the link.

The link is used most by shoppers who had not been to the store lately.
Among first arrivals at a store, people who had not visited it in the past week came through the link \lnkNewFirst\% of the time, against \lnkRecentFirst\% for recent visitors (+\lnkGap{} points \lnkGapCI).
After a specific-product answer, the link carries \lnkExtraNew\% of the extra visits from shoppers with no visit in the past week and \lnkExtraRecent\% from recent visitors (Figure~\ref{fig:routes}).

When a search comes in the 30 minutes before the first store visit, it names the answer's product \qArrPost\% of the time, against \qArrPlac\% in the earlier weeks.
The trip is quick: in the main group, half of first visits after a specific-product answer come within \medianMinSpec{} minutes.
Total searching shows no detectable drop: people run about as many searches that day as usual (\searchDay\% \searchDayCI, an interval that excludes a drop of more than \searchDayLoAbs\%).

\subsection{Then they keep browsing}\label{sec:store}
Once on the watch's page, our shopper looked at another watch.
The named product is often where the visit starts: after specific-product answers, a named product is the first page of the visit \rtpNamedLand\% of the time, other products \rtpOtherLand\%, and most of those landings follow a web search. Deeper in the visit, the store's own search and category pages lead to the named product about as often as to other products (\rtpNamedNav\% against \rtpOtherNav\% of later product views, at the stores where it can be identified).
The answer shapes which product people look at; the store's own pages guide the rest of the visit.
Against the same shopper's other visits, from every route and with link clicks left out, these visits open on a product page \allPdpWith\% of the time against \allPdpBase\% (Figure~\ref{fig:main}b; earlier period +\repAllPdp{} \repAllPdpCI).
We compare against all their visits because the route strongly affects where a visit opens: averaged within shoppers, typed visits rarely open on a product page (\ssPdpBase\%; \rtTypedPdp\% as a raw share in Figure~\ref{fig:routes}), generic searches often do (\genericPdpBase\%).

Shoppers keep browsing.
They view two or more products in \allTwoWith\% of visits against \allTwoBase\% on other visits (+\allTwo{} points \allTwoCI; earlier period +\repAllTwo{} \repAllTwoCI), the share of visits with at least two products the answer did not name shows no detectable change (\nvoOtherGap{} points \nvoOtherGapCI), and at the stores where it can be identified they view the store's own recommendations about as often as usual (\allRecWith\% against \allRecBase\%; a little more in the earlier period, +\repAllRec{} \repAllRecCI).
Counted over all specific-product answers, including those with no store visit, a view of a product the answer did not name follows \unnPlac{} answers in 100 in the earlier weeks and between \unnPostLo{} and \unnPostHi{} after the answer (at least +\unnLo{} points \unnLoCI; earlier period at least +\repUnnLo{} \repUnnLoCI; Figure~\ref{fig:main}a), so the higher store-visit rate comes with more browsing of other products, not only the named one. The figures are bounds because the match label says whether a day's product pages include the named product, not how many do.

\section{Discussion}
\paragraph{Referral reports miss these visits.} An answer names a product, and within minutes the shopper goes to look at it. Most show no sign of the assistant's link; many search for the product instead or go straight to the store, so the visit is logged as ordinary search or direct traffic.

Back to the running watch. Trailmere Sports records a search visit to the Contrail Run 5; the journey began earlier, with the assistant's answer. That gap between where a decision is made and where it is recorded is what referral reports cannot see, as with branded search ads~\cite{blake2015}.

\infobox{A boxed list of four actions for retailers, each tied to one finding.}{\textbf{What a retailer can do.} A referral report records the route into the store; the recommendation may have happened earlier.
\begin{itemize}\setlength{\itemsep}{1pt}
\item \emph{\rtUnlinked\% show no sign of the assistant's link.} Report assistant referrals next to product-page arrivals from search and direct traffic, and track product-name queries in search-console data. Link-only reporting captures about one in nine visits after a recommending answer, and about one in five of the extra visits.
\item \emph{Searches name the product.} Make product pages easy to find by the product names assistants recommend, in titles, addresses and site search.
\item \emph{Recent visitors rarely click.} Read the assistant's referral line as a signal about shoppers absent in the past week; its influence on recent visitors arrives by other routes.
\item \emph{Shoppers keep browsing.} Keep related products and recommendations within reach of the product page.
\end{itemize}}

\paragraph{Evaluating assistants and agents.}
Product follow-through gives a behavioral measure of a recommendation, without referral links or provider-side logs.
It can be reported separately for each assistant, as above, and extended to shopping agents.

\paragraph{For builders of shopping agents.}
People go to look at what the answer named, so the product an agent names, and what it says about it, shape which product people go to see.
And the answer is not the end of the journey: shoppers still browse other products at the store, so an agent that hands over a clear product name, one a search engine and a store both recognize, makes that step easy.

\paragraph{Open questions.} Telling whether these visits lead to purchases requires order data from the named stores. The causal test is to randomize which of several suitable products an assistant names and measure the visits that follow. And when agents buy on a person's behalf, the shopper may never open the store, so follow-through will have to be measured on the agent's purchases rather than on page views.

\paragraph{Limitations.}
This is an observational study. We read the named-product and store-visit rises as effects of the answer under one assumption: without the answer, a person's visits would have followed their earlier weeks and their equally active moments with no answer. Intent formed before asking could break it, though named products had been viewed in the prior week no more often than products visited on ordinary days (\pbNamedWeek\% vs.\ \pbOrdWeek\%; \repPbNamedWeek\% vs.\ \repPbOrdWeek\% earlier). The at-store and route results are descriptive.
We see visits, not purchases.
We chose the main group after first results; its store-visit rise (+\popHeadSpec{} points) is smaller than over all answers in which the assistant named the store (+\popPoolSpec).
The study covers two markets, twenty retailers and February to October 2026.

\section*{Ethics Statement}
This is secondary analysis of opt-in, de-identified browsing, search and assistant-conversation data collected under the panel provider's terms; no participant was contacted and no new data were collected.
Conversation text and visited pages were processed by third-party models as those terms allow.
Results are reported only in aggregate, and the linkage of conversations to browsing is used only for aggregate measurement, never to profile individuals.
AI coding assistants helped write the analysis and figure code and edit the text. The authors specified the design, verified every estimate and citation, and take full responsibility for all claims.

\bibliographystyle{ACM-Reference-Format}
\bibliography{references}

\end{document}